\pdfoutput=1

\documentclass[preprint,5p,twocolumn]{elsarticle}
\usepackage[utf8]{inputenc}
\usepackage[T1]{fontenc}
\usepackage[colorlinks=true
 ,urlcolor=blue
 ,anchorcolor=blue
 ,citecolor=red
 ,filecolor=blue
 ,linkcolor=blue
 ,menucolor=blue
 ,linktocpage=true
 ,pdfproducer=medialab
 ,pdfa=true
]{hyperref}
\journal{Physics Letter B}
\usepackage{amsmath,amsfonts,slashed,ulem}
\usepackage{cleveref}
\biboptions{sort&compress}
\usepackage{multicol}
\allowdisplaybreaks

\newcommand{\I}{\ensuremath{{\mathrm{i}}}}
\newcommand{\D}{\ensuremath{\mathrm{d}}}
\newcommand{\Nf}{\ensuremath{N_{\hspace{-0.7mm}f}}}
\newcommand{\Nc}{\ensuremath{N_{\hspace{-0.4mm}c}}}

\newcommand{\cL}{\ensuremath{\mathcal{L}}}
\newcommand{\eps}{\ensuremath{\varepsilon}}
\def\Tr{\mbox{Tr}}
\def\Ord{{\rm O}}
\def\be{\begin{equation}}
\def\ee{\end{equation}}
\newcommand{\xib}{\bar{\xi}}
\newcommand{\EP}{\mathcal{E}}
\newcommand{\Lpt}{\tilde{L}}
\newcommand{\eg}{{\it e.g.}}
\newcommand{\ie}{{\it i.e.}}
\AtBeginDocument{%
\setlength\abovedisplayskip{5pt}
\setlength\belowdisplayskip{5pt}}

\begin{document}

\thinmuskip=1mu
\medmuskip=2mu
\thickmuskip=3mu

%\begin{fmffile}{fmfsqed3m}
%\include{fig1.tex}

\begin{frontmatter}

\title{Electron mass anomalous dimension at $O(1/\Nf^2)$ \\ in three-dimensional $\mathcal{N}=1$ supersymmetric QED}

\author{S.~Metayer$^{\text{a}}$ and S.~Teber$^{\text{a}}$ \\
\vspace{0.25cm}
\it \small $^{\text{a}}$Sorbonne Universit\'e, CNRS, Laboratoire de Physique Th\'eorique et Hautes Energies, LPTHE, F-75005 Paris France
\vspace{-0.25cm}}
%\ead{smetayer@lpthe.jussieu.fr}
%\author[inst1]{}
%\ead{teber@lpthe.jussieu.fr}

%\affiliation[inst1]{organization={Sorbonne Universit\'e, CNRS, Laboratoire de Physique Th\'eorique et Hautes Energies, LPTHE, F-75005 Paris},country={France}}

\begin{abstract}
We consider massless three-dimensional $\mathcal{N}=1$ supersymmetric quantum electrodynamics (QED) with $\Nf$ flavours of electrons. Within the dimensional reduction scheme, we compute the critical exponents corresponding to both the electron and selectron field and (parity-even) mass anomalous dimensions at the next-to-leading order in the $1/\Nf$ expansion and in an arbitrary covariant gauge.
The equality of the gauge-invariant mass anomalous dimensions of the electron and the selectron is found to result from a subtle role played by the epsilon-scalars. 
Our general framework also allows us to compute the critical exponents of pure scalar QED and to recover known results in the case of spinor QED.
An application of our results to dynamical (s)electron mass generation is considered. We find evidence that, while dynamical flavor symmetry breaking occurs in spinor QED, both pure scalar QED and supersymmetric QED remain in an interacting conformal phase.

\end{abstract}

%\item 
%\item 
%\item 

% Keywords
%\begin{keywords}
% \sep \sep \sep
%\end{keywords}

\end{frontmatter}

\section{Introduction}
\label{sec:Introduction}
Three-dimensional massless quantum electrodynamics (QED$_3$) is a gauge field theory model of strongly interacting relativistic planar fermions that has been attracting continuous interest for the past four decades. Early studies \cite{Appelquist:1981vg,Appelquist:1981sf,Appelquist:1986fd} realized that, within a $1/\Nf$ expansion (where $\Nf$ is the number of electron flavours), this super-renormalizable model acquires an interacting fixed point in the infra-red (IR) and becomes effectively renormalizable. Large $\Nf$ techniques (see \cite{Gracey:2018ame} for a review) allow access to the critical properties of the model and in particular the field and mass anomalous dimensions that encode the renormalization of the composite operator $\bar{\psi} \psi$ \cite{Gracey:1993iu,Gracey:1993sn}. Such quantities play a crucial role in the study of fundamental quantum field theory mechanisms such as, \eg, dynamical symmetry breaking and electron mass generation~\cite{Pisarski:1984dj,Appelquist:1988sr}
(see recent progress in \cite{Gusynin:2016som,Kotikov:2016prf,Kotikov:2020slw,Metayer:2021rco}). In the last three decades, considerable interest in QED$_3$ also came from its applications to condensed matter physics systems with relativistic-like gapless quasiparticle excitations at low-energies such as high-$T_c$ superconductors
\cite{Dorey:1991kp,Franz:2001zz,Herbut:2002yq}, planar antiferromagnets
\cite{Farakos:1997hg} and graphene \cite{Semenoff:1984dq,Novoselov:2004}. 

Variants of QED$_3$ have also attracted continuous interest through the decades and in particular supersymmetric extensions of the model that will be of interest to us in the following. The case of $\mathcal{N}=2$ SQED$_3$ has been studied in the early paper of Pisarski \cite{Pisarski:1984dj} by dimensional reduction from the case of $\mathcal{N}=1$ four-dimensional supersymmetric QED (SQED$_4$) with focus on dynamical electron mass generation along the lines of the non-supersymmetric case. Actually, in $\mathcal{N}=1$ SQED$_4$, a non-perturbative non-renormalization theorem forbids dynamical mass generation \cite{Clark:1988zw} and it was therefore argued in \cite{Koopmans:1989kv} that this applies to $\mathcal{N}=2$ SQED$_3$, see also \cite{Walker:1999rs,CampbellSmith:1999vt} for further evidence. No such non-renormalization theorem holds in the case of $\mathcal{N}=1$ SQED$_3$ and there is no rigorous statement for electron mass generation in this case \cite{Koopmans:1989kv,CampbellSmith:1999aw}.
Nevertheless, this model has attracted significant attention recently in the context of the study of IR dualities and renormalization-group flows, see \cite{Khachatryan:2019yzg} and references therein for a review, and boundary field theories including models of super-graphene (in both the $\mathcal{N}=1$ and $\mathcal{N}=2$ cases) introduced in \cite{Herzog:2018lqz} (see also \cite{Gupta:2019qlg} for related non-perturbative computations of transport properties in the $\mathcal{N}=2$ case). Finally, within the condensed matter physics context, a potential emergence of supersymmetry in the low-energy limit of several lattice models was discussed during the last years, see, \eg, \cite{Lee:2006if,2013PhRvB87d1401R,Jian:2016zll,Zhao:2017bhw}. Computing critical exponents in the corresponding models is certainly valuable in order to assess the potential impact of supersymmetry on experimentally measurable observables. 

In the present paper, we will focus on massless three-dimensional $\mathcal{N}=1$ (minimal) SQED$_3$ with $\Nf$ two-component Dirac fermions. Along the lines of the non-supersymmetric case, we will compute exactly the critical exponents of the model at the next-to-leading order (NLO) in the large $\Nf$ expansion and in an arbitrary covariant gauge. Our computations will be carried out in the component formalism within the dimensional reduction (DRED) scheme \cite{Siegel:1979wq,Siegel:1980qs,Capper:1979ns} (see also \cite{Jack:1997sr} for a review) that is the most convenient regularization scheme for practical calculations in supersymmetric theories. We will also carry out our calculations with sufficient generality in order to highlight the features brought by supersymmetry (SUSY). In particular, we will find that the electron and selectron critical exponents are highly constrained and that the related identities (by analogy with four-dimensional supersymmetric Slavnov-Taylor identities \cite{Hollik:1999xh,Rupp:2000vi}) are achieved thanks to a subtle role played by the so called epsilon-scalars. Our general framework will also allow us to consider the sub-case of bosonic (or pure\footnote{Here ``pure'' means no $\phi^4$-type interactions.} scalar) QED$_3$ (bQED$_3$) and derive the critical exponents at NLO for this model too. And the sub-case of fermionic (or spinor) QED$_3$ (fQED$_3$) for which we shall recover known results as a useful check of our calculations. Lastly, we will always assume that the number of flavours $\Nf$ is even. This implies that we shall focus on the parity-even mass anomalous dimension (it has been proved that parity-odd masses cannot be dynamically generated \cite{Vafa:1984xg,Appelquist:1986qw}).

The paper is organized as follows. In sec.\ \ref{sec:Model}, we present the model and conventions that will allow us to compute its critical exponents within the DRED scheme. In sec.\ \ref{sec:setup}, we set up the $1/\Nf$ expansion and provide some technical details underlying our calculations. The field anomalous dimensions of SQED$_3$, bQED$_3$ and, as a check, fQED$_3$ at the NLO of the $1/\Nf$ expansion are then presented in sec.\ \ref{sec:field} and the corresponding mass anomalous dimensions in sec.\ \ref{sec:mass}. 
An application of our results to dynamical (s)electron mass generation is discussed in sec.\ \ref{sec:massgen}. We summarize our results and conclude in sec.\ \ref{sec:conclusion}. For completeness, the renormalized self-energies of SQED$_3$, bQED$_3$ and fQED$_3$ are given in app.\ \hyperref[sec:app1]{A}.

\section{Model and conventions}
\label{sec:Model}
In the DRED scheme, we assume that coordinates are $d$-dimensional (with $d=3-2\eps$) in order to
regulate the divergent Feynman integrals while fields remain three-dimensional as required by SUSY.
Within this scheme, the Lagrangian of massless $\mathcal{N}=1$ supersymmetric QED$_3$ reads
\begin{align}
\cL & = \I \bar{\psi}\slashed{D}\psi-\frac{1}{4} \hat{F}_{\mu\nu}^2-\frac{1}{2\xi}(\hat\partial_{\mu} \hat{A}^{\mu})^2\nonumber \\
& +\frac{\I}{2}\bar{\lambda} \slashed{\partial}\lambda + |\hat{D}_{\mu} \phi|^2 - \I e(\bar{\psi} \lambda \phi - \bar{\lambda} \psi \phi^\ast)+|F|^2 \nonumber \\
& -\frac{1}{2} (\hat\partial_{\mu} \bar{A}_{\nu})^2
- e \bar{\psi}\,\,\bar{\gamma}^{\mu}\bar{A}_{\mu}\psi 
+ e^2 \bar{A}^2 \lvert\phi\rvert^2 \,,
\label{eq:action}
\end{align}
where $\hat{D}_\mu = \hat{\partial}_\mu + \I e \hat{A}_\mu$ and $\slashed{t}=\hat\gamma^\mu \hat{t}_\mu$ for any $d$-vector $t$.\footnote{All computations will be carried out considering that, in three dimensions, odd (hatted) gamma matrix traces are non-vanishing, \ie, $\Tr(\hat{\gamma}^{\mu}\hat{\gamma}^{\nu}\hat{\gamma}^{\rho})\propto \hat{\eps}^{\mu\nu\rho}$, with, \eg, $\hat{\gamma}^\mu=\{\sigma_2,\I\sigma_1,\I\sigma_3\}$ and $\sigma_i$ the Pauli matrices. As expected from a parity-even theory, we find that such terms never contribute to any result.\label{footn:ec3}}
The Lagrangian of eq.\ (\ref{eq:action}) is built from $\Nf$ matter multiplets $\{ \psi, \phi, F\}$, where $\psi$ is a $2$-component Dirac 
fermion (electron), $\phi$ a complex pseudo-scalar (selectron) and $F$ a complex auxiliary scalar field, as well as a gauge multiplet $\{\hat{A}_{\mu},\bar{A}_{\mu},\lambda\}$, where $\hat{A}_{\mu}$ is the $d$-dimensional $U(1)$ gauge field (photons), $\bar{A}_{\mu}$ is the $2\eps$-dimensional $U(1)$ gauge field ($\eps$-scalar), and $\lambda$ a $2$-component Majorana fermion (photino). Here we use the notations of the review \cite{Mihaila:2013wma} where hatted (respectively barred) quantities 
have $d$ (respectively $3-d$) components. Additionally, in (\ref{eq:action}), $\xi$ is the covariant gauge fixing\footnote{The gauge fixing term is SUSY breaking, therefore only the physical gauge-invariant quantities will be SUSY invariant.} parameter and $e^2$ is the coupling constant of the theory with dimension of mass. 

In order to highlight SUSY effects in our computations, each superpartner field will be associated with a factor $S\in\{0,1\}$ such that
\be
\Phi\rightarrow S\,\Phi \quad \forall\,\Phi\in\{\bar{A}^{\mu},\,\lambda,\,\phi\}\,,
\ee
and $S^2=S$. Hence, at any step of the calculation we may turn on (respectively off) SUSY by setting $S=1$ (respectively $S=0$). This allows us to check our expressions by recovering known results for corresponding non-SUSY theories such as large-$\Nf$ fQED$_3$ \cite{Gracey:1993iu,Gracey:1993sn}, see also, \eg, \cite{Metayer:2021rco} and references therein. Because the latter are generally expressed in terms of $4$-component spinors, instead of $2$-component spinors, we shall work with $n$-component ones. Moreover, the case $n=0$ corresponds to bQED$_3$. In order to keep track of all of these cases while limiting the complexity of our formulas, we shall therefore impose the constraint $n(n-2)S=0$.

Similarly, to better appreciate the effects of DRED during the computations, the $\eps$-scalar field will be associated with a factor $\EP \in\{0,1\}$ such that
\be
\bar{A}^{\mu}\rightarrow \EP \,\bar{A}^{\mu}\,,
\ee
and $\EP ^2=\EP $. Indeed, as we shall see in the following, though $\eps$-scalars affect only few quantities at NLO, their effect is crucial to ensure the validity of supersymmetric identities. From our general formulas, the case of minimal SQED$_3$ corresponds to $S=\EP =1$ and $n=2$. The sub-case of fQED$_3$ will be recovered with the help of $S=\EP =0$ and $n=2$ and that of bQED$_3$ with the help of $S=1$ and $n=\EP=0$.

With these notations defined, the dressed Feynman propagators (in Minkowski space) associated with (massless) $\mathcal{N}=1$ SQED$_3$ read
\begin{subequations}
\label{gmp+mmp}
\begin{flalign}
& \hspace{2.25cm}\hat{G}_{AA}^{\mu \nu} (p) =\frac{-\I}{1 - \Pi^\gamma(p^2)} \frac{\hat{d}^{\mu\nu}}{p^2}\,,
\label{gmp:AAd} \hspace{-2.25cm}&& \\
&\bar{G}_{AA}^{\mu \nu}(p)=\frac{-\I \EP S}{1 - \Pi^\eps(p^2)} \frac{\bar{g}^{\,\mu\nu}}{p^2} \,, 
&& G_{\lambda {\bar{\lambda}}}\hspace{0.08cm}(p)= \frac{\I S}{1 - \Pi^\lambda(p^2)}\frac{1}{\slashed{p}}\,,
\label{gmp:AAeps+ll}\\
&G_{\psi {\bar{\psi}}}\hspace{0.05cm}(p) = \frac{\I}{1-\Sigma_p^\psi(p^2)}\frac{1}{\slashed{p}}\,,
&& G_{\phi {\phi^\dagger}}\hspace{-0.05cm}(p) = \frac{\I S}{1-\Sigma_p^\phi(p^2)}\frac{1}{p^2}\,,
\label{mmp:psipsi+phiphi}
\end{flalign}
\end{subequations}
where $\hat{d}^{\mu\nu}={\hat{g}^{{\mu}{\nu}} - (1 - \xi)\, \hat p^{\mu} \hat p^{\nu}/p^2}$, $p^2=\hat p^\mu \hat p_\mu$
and the photon propagator (\ref{gmp:AAd}) is expressed in a non-local $\xi$-gauge \cite{Nash:1989xx,Kugo:1992pr,Simmons:1990bz}. The gauge multiplet self-energies in (\ref{gmp+mmp}) are parameterized as
\begin{subequations}
\label{eq:polarizations}
\begin{align}
&\hat{\Pi}^{\mu \nu}(p)=(p^2 \hat{g}^{\,\mu \nu} - \hat p^{\mu} \hat p^{\nu})\Pi^\gamma(p^2)\,,\\
&\bar{\Pi}^{\mu \nu}(p)= p^2 \bar{g}^{\,\mu \nu} \Pi^\eps(p^2)\,,\\
&\Pi^{\lambda\phantom{a}}(p)= \slashed{p}\,\Pi^\lambda(p^2)\,,
\end{align}
\end{subequations}
and the matter multiplet self-energies (which will be our main focus in this paper) as
\begin{subequations}
\label{eq:matter-SE}
\begin{align}
\Sigma^\psi(p) & = \slashed{p}\hspace{0.14cm}\Sigma_p^\psi(p^2) + m_\psi\Sigma_m^\psi(p^2)\,, \\
\Sigma^\phi(p) & = p^2\Sigma_p^\phi(p^2)+ m_\phi^2\hspace{0.05cm}\Sigma_m^\phi(p^2)\,,
\end{align}
\end{subequations}
where the masses $m_x$ have been introduced as IR regulators $m_x \ll p$, with $x=\psi,\phi$. 
Hence, after extracting $\Sigma_p$ and $\Sigma_m$ for both electrons and selectrons, the masses $m_x$ will be sent back to zero. Our computations will therefore all be efficiently carried out with the massless Feynman diagram techniques, see, \eg, the review \cite{Kotikov:2018wxe}.

As for our renormalization scheme, we shall work in the modified minimal reduction ($\overline{\text{DR}}$) scheme that subtracts the Euler constant $\gamma_E$ as well as a factor of $4\pi$ from the $\eps$ expansion, \ie, $\overline{\mu}^2 = 4\pi e^{-\gamma_E}\mu^2$ where $\mu$ is the renormalization scale in the DRED scheme such that $\Nf$ trivially renormalize as $\Nf\rightarrow\mu^{-2\eps}\Nf$. Indeed, we recall that, in the large $\Nf$ limit, SQED$_3$ \cite{Koopmans:1989kv}, similarly to bQED$_3$ \cite{Appelquist:1981vg,Appelquist:1981sf} and fQED$_3$ \cite{Pisarski:1984dj,Appelquist:1986fd}, is a non-running (``standing'') gauge theory, \ie, the coupling is not renormalized, implying finite polarizations (\ref{eq:polarizations}) and therefore vanishing beta functions.\footnote{\label{footn:ec1} These include the beta function for the effective coupling, $\overline{\alpha}(k) = (\overline{\alpha}/k)/(1-\Pi^\gamma(k^2))$ where $\overline{\alpha}=e^2 \Nf$, that is particularly appropriate in a $1/\Nf$-expansion and reads: $\beta(\overline{\alpha})=-\overline{\alpha}(1-\overline{\alpha}/\overline{\alpha}^*)$ where $\overline{\alpha} \rightarrow 0$ at the asymptotically free UV fixed point while $\overline{\alpha} \rightarrow \overline{\alpha}^*$ at the interacting IR fixed point, see \cite{Appelquist:1981vg,Appelquist:1981sf,Appelquist:1986fd} as well as the more recent \cite{Gusynin:2020cra} and also footnote \ref{footn:ec2}.} This leads to the triviality of the renormalization constants for the coupling, gauge-multiplet fields and gauge-fixing parameter, formally $Z_x=1$ with $x\in\{e,\,\hat{A}^{\mu},\,\bar{A}^{\mu},\,\lambda,\,\xi\}$. 
The remaining non-trivial renormalization constants are therefore for matter fields and masses, with conventions
\begin{subequations}
\label{eq:Zdef}
\begin{align}
& \psi=Z_\psi^{1/2}\psi_r\,, && \phi=Z_\phi^{1/2}\phi_r\,, \label{eq:Zdefa}\\
& m_\psi=Z_{m_\psi}m_{\psi\,r}\,, && m_\phi=Z_{m_\phi}m_{\phi\,r}\,.
\label{eq:Zdefb}
\end{align}
\end{subequations}
These can be computed from the self-energies (\ref{eq:matter-SE}) via the relations
\begin{subequations}
\label{eq:Zrel}
\begin{align}
\left(1-\Sigma_p^\psi\right)Z_\psi=\text{finite}\,, && \left(1-\Sigma_p^\phi\right)Z_\phi=\text{finite}\,,
\label{eq:Zrela}\\
\frac{1+\Sigma_m^\psi}{1-\Sigma_p^\psi}Z_{m_\psi}=\text{finite}\,, && \frac{1+\Sigma_m^\phi}{1-\Sigma_p^\phi}Z^2_{m_\phi}=\text{finite}\,, 
\label{eq:Zrelb}
\end{align}
\end{subequations}
where ``finite'' means of the order of $\eps^0$. Finally, the associated anomalous dimensions are defined as
\be
\gamma_x=\frac{\D\log Z_x}{\D\log \mu}\,,~~~x\in\{\psi,\,\phi,\,m_\psi,\,m_\phi\}\,,
\label{eq:gammadef}
\ee
and correspond to the critical exponents we wish to compute in a $1/\Nf$ expansion at the non-trivial IR fixed point. 

\section{Setting up the \texorpdfstring{$1/\Nf$}{} expansion}
\label{sec:setup}

The $1/\Nf$ expansion amounts to sum an infinite chain of simple matter loops in force field propagators, \ie, the gauge multiplet propagators of eqs.\ (\ref{gmp:AAd}) and (\ref{gmp:AAeps+ll}), making the theory effectively renormalizable in the IR \cite{Appelquist:1986fd}. Up to NLO, the gauge multiplet polarizations, first obtained in \cite{James:2021ggq} for SQED$_3$ and conveniently generalized to our framework, take the form
\be
\Pi^x(p^2) = \Pi_1^x(p^2)\,\bigg[ 1 + \frac{C_x}{\Nf} + \Ord\big(1/\Nf^2\big) \bigg] \,,
\label{Pix:gen}
\ee
with $x\in\{\gamma,\,\eps,\,\lambda\}$. In (\ref{Pix:gen}), the leading order (LO) contributions read
\be
\Pi^{\gamma}_1=-\frac{{(n+2S)}a}{2p}\,,
~~~\Pi^{\eps}_1=-\frac{2\EP Sa}{p}\,,
~~~\Pi^{\lambda}_1=-\frac{2Sa}{p}\,,
\label{Pi:LO}
\ee
with the Euclidean momentum $p=\sqrt{-p^2}$, as well as $a=\Nf e^2/16$ and the NLO coefficients
\begin{subequations}
\label{eq:Cfactors}
\begin{align}
& C_\gamma={\frac{8n(92-9\pi^2)}{9(n+2S)^2\pi^2}+\frac{16(164-20n-9\pi^2)S}{9(n+2)^2\pi^2}}\,, \label{eq:Cgamma}\\
& C_\eps=\frac{2(12-\pi^2)S}{\pi^2}\,, \quad C_\lambda=\frac{2(38-2\EP -3\pi^2)S}{3\pi^2}\, . \label{eq:Ceps+Cll}
\end{align}
\end{subequations}
In the non-SUSY ($S=0$) case, the only non-zero coefficient is $C_\gamma^{(\text{f})}=2(92-9\pi^2)/9\pi^2$ for 4-component electrons \cite{Gusynin:2000zb,Teber:2012de,Kotikov:2013kcl}. 
In the SUSY ($S=1$) case, all polarization corrections (\ref{eq:Cfactors}) are equal, provided that $\eps$-scalars are allowed ($\EP =1$) and read $C_\gamma=C_\eps=C_\lambda=2(12-\pi^2)/\pi^2$ for 2-component spinors.
In the bosonic ($n=0$) case, the coefficient reads $C_\gamma^{(\text{b})}=4(164-9\pi^2)/9\pi^2$. 
\begin{table}[t]
\centering
\begin{tabular}{l|l}
{fQED$_3$ ($n=2$)} & $\tilde\Pi^\gamma=1 + 0.1429/\Nf +\Ord(1/\Nf^2)$ \\
{SQED$_3$ ($n=2$)} & $\tilde\Pi^\gamma=2 + 0.8634/\Nf +\Ord(1/\Nf^2)$ \\
{bQED$_3$ ($n=0$)} & {$\tilde\Pi^\gamma=1 + 3.3852/\Nf +\Ord(1/\Nf^2)$}
\end{tabular}
\vspace*{-0.25cm}
\caption{Numerical polarizations $\tilde\Pi^\gamma=-p\, \Pi^\gamma/a$}
\vspace*{-0.25cm}
\label{tab:pinum}
\end{table}
In order to further appreciate the differences between the various models, we provide comparative numerical results in tab.\ \ref{tab:pinum}.
Note that in contrast to fQED$_3$ radiative corrections rather strongly affect the bQED$_3$ polarization.\footnote{\label{footn:ec2} Following footnote \ref{footn:ec1}, the coupling at the non-trivial IR fixed point $\overline{\alpha}^*$ is given by: $\overline{\alpha}^*=16/\tilde\Pi^\gamma = 16/(1+C_\gamma/N_f)$ (see eq.~(3.3) in \cite{Appelquist:1981sf} where $A$ in that paper corresponds to our $C_\gamma$ for either fQED$_3$ or bQED$_3$). In the case of fQED$_3$ radiative corrections weakly affect the fixed point ($C_\gamma=0.1429$ for $2$-component spinors) while the effect is much more pronounced for bQED$_3$ ($C_\gamma=3.3852$).} The contribution
of the scalar (bosonic) field also enhances the SQED$_3$ polarization, a fact first noticed at LO in \cite{Koopmans:1989kv}; they also affect the NLO SQED$_3$ polarization though not as much as in bQED$_3$ as their effect is tempered by the contribution of the fermions.

Substituting (\ref{Pix:gen}) in (\ref{gmp+mmp}) and keeping $a$ fixed as $\Nf$ goes to infinity while focusing on the IR limit $p \ll a$, we deduce the LO gauge field propagators 
\begin{subequations}
\label{gmp:1/Nf}
\begin{align}
&\hspace{1.5cm}\hat{G}_{AA}^{\mu \nu}(p) = {\frac{2\I}{(n+2S)a}}\frac{\hat{d}^{\mu\nu}}{p}\,,
\label{gmp:photonprop}\\
& \bar{G}_{AA}^{\mu \nu}(p) = \frac{\I\EP S}{2a}\frac{\bar{g}^{\,{\mu}{\nu}}}{p}\,, 
\qquad G_{\lambda {\bar{\lambda}}}(p) =\frac{-\I S}{2a}\frac{\slashed{p}}{p}\, .
\end{align}
\end{subequations}
Our Feynman rules therefore consist of (\ref{gmp:1/Nf}) together with the matter propagators  (\ref{mmp:psipsi+phiphi}) at LO
\begin{equation}
G_{\psi {\bar{\psi}}}\hspace{0.05cm}(p) = \frac{\I}{\slashed{p}},
\qquad
G_{\phi {\phi^\dagger}}\hspace{-0.05cm}(p) = \frac{\I S}{p^2}\,,
\end{equation}
and the vertices
\begin{align}
& \hat{\Gamma}^{\mu}_{A\psi\bar\psi}=-\I e \hat{\gamma}^{\mu}\,, 
&& \hspace{-0.25cm} \bar{\Gamma}^{\mu}_{A\psi\bar\psi}=-\I e \EP S\,\bar{\gamma}^{\mu}\,, \nonumber \\
& \hat{\Gamma}^{\mu \nu}_{AA\phi\phi^\dagger}=2\I e^2 S \hat{g}^{\mu \nu}\,, 
&& \hspace{-0.25cm} \bar{\Gamma}^{\mu \nu}_{AA\phi\phi^\dagger}=2\I e^2 \EP S\, \bar{g}^{\mu \nu} \nonumber\\
& \Gamma_{\bar\lambda\psi\phi}=eS\,, 
&& \hspace{-0.25cm} \Gamma_{\lambda\bar\psi\phi}=-eS\,, \nonumber\\
& \hat{\Gamma}^{\mu}_{A\phi\phi^\dagger}=-\I e S (\hat{p}+\hat{k})^{\mu} \, .
&&
\end{align}
From this combination of propagators and vertices, the mass scale $e^2$ drops out of self-energies in favor of the dimensionless coupling constant $1/\Nf$. Diagrams are then generated according to their power of $1/\Nf$ with each matter loop bringing a factor of $\Nf$ and an additional minus sign from each fermionic loop.\footnote{Because this model involves both Dirac and Majorana fermions, the fermionic flows have been carefully treated using the conventions of \cite{Denner:1992me,Denner:1992vza}.}

\begin{figure}[t!]
\centering
%\GRAPHLOAeps ~~ \GRAPHLOAAA ~~ \GRAPHLOAepseps
\includegraphics[scale=1.0]{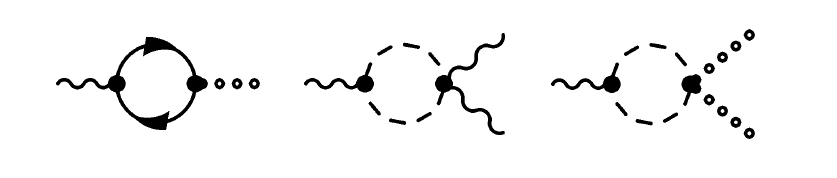}
\vspace*{-0.25cm}
\caption{Exactly vanishing matter-loop diagrams (dotted line for the $\eps$-scalar, wavy line for the photon, solid line with an arrow for the electron and dashed line for the selectron).}
\vspace*{-0.25cm}
\label{fig:1}
\end{figure}

Given the variety of propagators and vertices in SQED$_3$, a multitude of diagrams is a priori generated at each order of the large $\Nf$ expansion. Interestingly, a number of LO matter loop diagrams are found to vanish. The first vanishing diagrams are those of Furry kind, with loops composed of an odd number of legs and matter propagators. Indeed, at LO, %the 8 
we find that all the possible three-legged matter triangle diagrams vanish, either exactly or pairwise with opposite matter flows. Moreover, three additional LO diagrams of matter-bubble type, presented in fig.\ \ref{fig:1}, vanish exactly. 
The first diagram in fig.\ \ref{fig:1} vanishes as it involves a trace over an odd number of barred gamma matrices, \eg, $\Tr(\bar{\gamma}^{\mu})=0$.\footnote{We take the Levi-Civita symbol in $2\eps$-dimension to be vanishing, implying vanishing traces of odd numbers of barred gamma matrices, as opposed to the case of hatted gamma matrices, see footnote \ref{footn:ec3}.} The two other selectron bubble diagrams in fig.\ \ref{fig:1} are exactly zero by parity. Therefore, every diagram containing one of these sub-topologies can be discarded. This not only tremendously reduces the number of NLO diagrams, but also guarantees that matter-loops are connected by simple chains of force field propagators, in accordance with our starting assumption.

\section{Field anomalous dimensions}
\label{sec:field}

At NLO, the computation of the electron and selectron self-energies involve 17 and 14 distinct diagrams, respectively (by comparison, the corresponding non-SUSY NLO computations involve only 3 diagrams for the electron). 

In this section, we focus on the purely massless case and extract the functions $\Sigma_p^x(p^2)$ ($x=\{ \psi,\,\phi \}$) from eqs.\ (\ref{eq:matter-SE}). All calculations done, up to the NLO in the $1/\Nf$ expansion, they read
\begin{subequations}
\label{eq:Sigmap}
\begin{flalign}
& \Sigma^\psi_p =-\frac{2(S-\xib)}{R\Nf\eps}-\frac{2(S-\xib)^2}{R^2\Nf^2\eps^2}-\frac{1}{3R^2\Nf^2\eps}\bigg[4+(77+6\EP)S& \nonumber \\[-0.3cm] 
& \\[-0.3cm]
& \hspace{0cm}+4\big(1-(19+3\EP)S+6\xib\big)\xib-6R\big(SC_\lambda-\xib C_\gamma\big)\bigg]+\Ord(\eps^0)\,, & \nonumber\\[-0.1cm] 
& {\Sigma^\phi_p} =\frac{(6-n+2\xib)S}{R\Nf\eps}-\frac{(6-n+2\xib)^2S}{2R^2\Nf^2\eps^2}-\frac{S}{6R^2\Nf^2\eps}\bigg[8(85+28\xib) &\nonumber\\[-0.3cm]
& \\[-0.3cm]
& \hspace{0cm}-n(163+40\xib)-12\EP-6R\big(nC_\lambda-2(3+\xib)C_\gamma\big)\bigg]+\Ord(\eps^0)\,,& \nonumber
\end{flalign}
\end{subequations}
with 
$\bar{\xi}=(2-3\xi)/2$,
$R=A(-4p^2/\bar\mu^2)^{\eps}$ and $A=3\pi^2(n+2S)/8$.
Note that, in (\ref{eq:Sigmap}), $\eps$-scalars do contribute to the self-energies. Their contribution arises partly from the polarization correction $C_\lambda$ but not from $C_\eps$, see (\ref{eq:Ceps+Cll}). From (\ref{eq:Sigmap}), together with the definitions (\ref{eq:Zdefa}), (\ref{eq:Zrela}) and (\ref{eq:gammadef}), we deduce the following matter-field anomalous dimensions
\begin{subequations}
\label{eq:gammafields}
\begin{flalign}
\gamma_\psi& =\frac{4(S-\xib)}{A\Nf}+\frac{4}{3A^2\Nf^2}\bigg[4+(29-6\EP)S\label{eq:gammapsi}\\[-10pt] 
&\hspace{3.6cm}-3A\big(SC_\lambda-\xib C_\gamma\big)\bigg]+\Ord\big(1/\Nf^3\big)\,,\nonumber\\[-5pt]
{\gamma_\phi}& =\frac{2(n-6-2\xib)S}{A\Nf}+\frac{2S}{3A^2\Nf^2}\bigg[8+29n-12\EP\label{eq:gammaphi}\\[-7pt]
&\hspace{2.7cm}-3A\big(nC_\lambda-(8-3\xi)C_\gamma\big)\bigg]+\Ord\big(1/\Nf^3\big)\,.\nonumber
\end{flalign}
\end{subequations}
As a check on our results, eq.\ (\ref{eq:gammapsi}) allows us to recover the {fQED$_3$} electron field anomalous dimension, \ie, for $S=0$ and $n=4$-component spinors
\be
\gamma_\psi^{{(\text{f})}}=-\frac{8\xib}{3\pi^2\Nf}+\frac{16\big(4+(92-9\pi^2)\xib\big)}{27\pi^4\Nf^2}+\Ord(1/\Nf^3)\,,
\label{eq:gammapsinonsusy}
\ee
first obtained, with an other method, in the Landau gauge in \cite{Gracey:1993iu} and in an arbitrary covariant gauge in \cite{Gracey:1993sn}. In the supersymmetric ($S=1$) case, for $n=2$-component spinors, eqs.\ (\ref{eq:gammafields}) yield
\begin{subequations}
\label{eq:fieldanomalousdims}
\begin{flalign}
\gamma_\psi & =\frac{4\xi}{\pi^2\Nf}+\frac{8\big(2-(12-\pi^2)\xi\big)}{\pi^4\Nf^2}+\Ord(1/\Nf^3)\,, \hspace{-2cm} & \\
\gamma_\phi & =\frac{4(\xi-2)}{\pi^2\Nf}+\frac{8\big(26-2\pi^2-(12-\pi^2) \xi\big)}{\pi^4\Nf^2} +\Ord(1/\Nf^3)\,, \hspace{-2cm} & 
%\raisetag{34pt}
\end{flalign}
\end{subequations}
where $\gamma_\psi\neq\gamma_\phi$ possibly because our gauge fixing method breaks SUSY. Interestingly, (\ref{eq:fieldanomalousdims}) is $\EP$-independent up to NLO, so that $\eps$-scalars completely cancel out in the matter field anomalous dimensions.
Finally, in the case of bQED$_3$, the scalar field anomalous dimension can be computed from eq.\ (\ref{eq:gammaphi}) by taking $n=\EP=0$ and $S=1$, reading
\be
\label{eq:fieldanomalousdims:bQED}
\gamma_\phi^{{(\text{b})}}=-\frac{8 (8-3 \xi)}{3 \pi ^2 \Nf}+\frac{32(440-164\xi-3\pi^2(8-3\xi))}{9 \pi ^4 \Nf^2}+\Ord(1/\Nf^3)\,,
\ee
{where the LO contribution in the Landau gauge is in accordance with \cite{Khachatryan:2019veb} 
while the NLO one is a new result. Eqs.~(\ref{eq:fieldanomalousdims}) and (\ref{eq:fieldanomalousdims:bQED}) and are the first set of main results of this paper.

\section{Mass anomalous dimensions}
\label{sec:mass}

From the same set of diagrams and allowing for a small mass $m_x$ with ($x=\{ \psi,\,\phi \}$) for the (s)electron, we extract the functions $\Sigma_m^x(p^2)$ from eqs.\ (\ref{eq:matter-SE}). All calculations done, up to the NLO in the $1/\Nf$ expansion, they read
\begin{subequations}
\label{eq:Sigmam}
\begin{flalign}
& \Sigma^\psi_m = \frac{3(2+\xi)}{R\Nf\eps}+\frac{9(2+\xi)^2}{2R^2\Nf^2\eps^2}+\frac{1}{R^2\Nf^2\eps}\bigg[220-21S&\nonumber\\[-5pt]
& \hspace{0.9cm}-4(29-4\xib)\xib+3(2+\xi)\big(6\EP S-RC_\gamma\big)\bigg]+\Ord(\eps^0)\,,\hspace{-2cm}& \\[-5pt]
& {\Sigma^\phi_m} =\frac{3(n+\xi)S}{R\Nf\eps}+\frac{9(n+\xi)^2S}{2R^2\Nf^2\eps^2}+\frac{3S}{2R^2\Nf^2\eps}\bigg[81n&\nonumber\\[-5pt]
& \hspace{1.13cm}+12\EP-8(2+\xib)\xib-2R\big(nC_\lambda+\xi C_\gamma\big)\bigg]+\Ord(\eps^0)\,,\hspace{-2cm}&
\end{flalign}
\end{subequations}
with the same notations as for eqs.\ (\ref{eq:Sigmap}).
In (\ref{eq:Sigmam}), similarly to (\ref{eq:Sigmap}), 
we note that $\eps$-scalars contribute to the self-energies in part from the polarization correction 
$C_\lambda$ (this time for the selectron only) but not from $C_\eps$, see (\ref{eq:Ceps+Cll}).
From (\ref{eq:Sigmam}), together with the definitions (\ref{eq:Zdefb}), (\ref{eq:Zrelb}), (\ref{eq:gammadef}) and our previous 
results (\ref{eq:Sigmap}), we deduce the following mass anomalous dimensions
\begin{subequations}
\label{eq:res:inter:gammam}
\begin{flalign}
\gamma_{m_\psi}& =\frac{4(4-S)}{A\Nf}+\frac{8}{3A^2\Nf^2}\bigg[16-(46-3\EP)S\label{eq:gammampsi} \\[-10pt]
& \hspace{3.4cm}+\tfrac{3}{2}A(SC_\lambda-4C_\gamma)\bigg]+\Ord\big(1/\Nf^3\big)\,,\nonumber\\[-7pt] 
{\gamma_{m_\phi}}&=\frac{2(4+n)S}{A\Nf}-\frac{8S}{3A^2\Nf^2}\bigg[28-15\EP+7n\label{eq:gammamphi}\\[-10pt] 
& \hspace{3.4cm}+\tfrac{3}{4}A(nC_\lambda+4C_\gamma)\bigg]+\Ord\big(1/\Nf^3\big)\, . \nonumber
\end{flalign}
\end{subequations}
Reassuringly, eqs.\ (\ref{eq:res:inter:gammam}) are gauge invariant which is the first check on these results. Moreover, eq.\ (\ref{eq:gammampsi}) allows us to recover the {fQED$_3$} electron mass anomalous dimension, \ie, for $S=0$ and $n=4$-component spinors
\begin{flalign}
\label{eq:gammam:qed}
\gamma_{m_\psi}^{{(\text{f})}}=\frac{32}{3\pi^2\Nf} + \frac{64(3\pi^2-28)}{9\pi^4\Nf^2}+\Ord(1/\Nf^3)\,,
\end{flalign}
first obtained in \cite{Gracey:1993sn} (up to a conventional sign). 
In the supersymmetric ($S=1$) case, for $n=2$-component spinors, eqs.\ (\ref{eq:res:inter:gammam}) yield 
\begin{subequations}
\label{eq:gammawithEps}
\begin{align}
& \gamma_{m_\psi}=\frac{8}{\pi^2\Nf}-\frac{16(14-\pi^2)}{\pi^4\Nf^2}+\Ord(1/\Nf^3)\,,& \label{eq:gammawithEpsa}\\
& \gamma_{m_\phi}=\frac{8}{\pi^2\Nf}-\frac{16(46-4\EP -3\pi^2)}{3\pi^4\Nf^2}+\Ord(1/\Nf^3)\,.\hspace{-3cm}& \label{eq:gammawithEpsb}
\end{align}
\end{subequations} 
Interestingly, unlike for the field anomalous dimensions, $\eps$-scalars do contribute to (\ref{eq:gammawithEpsb}) at the NLO
of the $1/\Nf$ expansion. Actually, their effect is crucial because upon setting $\EP =1$ in eqs.\ (\ref{eq:gammawithEps}), we find that
\be
\gamma_{m_\psi}=\gamma_{m_\phi}=\frac{8}{\pi^2\Nf}-\frac{16(14-\pi^2)}{\pi^4\Nf^2}+\Ord(1/\Nf^3)\,,
\label{eq:res:gammam}
\ee
where the LO contribution agrees with \cite{Benvenuti:2018cwd} and the NLO one is new. Our analysis confirms that $\eps$-scalars are crucial to ensure the equality of the mass anomalous dimensions of the electron and the selectron up to NLO. As anticipated in the Introduction, this result is a perturbative proof that the identity $\gamma_{m_\psi}=\gamma_{m_\phi}$ holds in the three-dimensional context, by analogy with the four-dimensional supersymmetric Slavnov-Taylor identities. Finally, the bQED$_3$ case can be accessed with the help of eq.~(\ref{eq:gammamphi}) with $n=0$ (and $\EP=0$, $S=1$) which yields
\be
\label{eq:gammam:bqed}
\gamma_{m_\phi}^{{(\text{b})}}=\frac{32}{3\pi^2\Nf}-\frac{128(64-3\pi^2)}{9\pi^4\Nf^2}+\Ord(1/\Nf^3)\,,
\ee
where the LO agrees with \cite{Benvenuti:2018cwd} and the NLO one is a new result. 
Eqs.~(\ref{eq:res:gammam}) and (\ref{eq:gammam:bqed}) are the second set of main results of this paper.

In order to further appreciate the differences between the various models, we provide comparative numerical results in tab.\ \ref{tab:gammanum}. Comparing the cases of fQED$_3$ and bQED$_3$, we see that (similarly to the case of the polarization operator displayed in tab.~\ref{tab:pinum}), NLO radiative corrections are stronger in absolute value in the bosonic case and are affected by a negative sign. The case of SQED$_3$ is somehow intermediate between fQED$_3$ and bQED$_3$ with a tendency of the bosonic contribution from the scalar field to reduce the overall electron mass anomalous dimension, due to a smaller LO term and a negative NLO contribution.} This is to be contrasted with the enhancement of the SQED$_3$ photon polarization displayed in tab.~\ref{tab:pinum}.
\begin{table}[t!]
\centering
\begin{tabular}{l|l}
{fQED$_3$ ($n=2$)} & $\gamma_{m_\psi}=2.162/\Nf + 0.470/\Nf^2+\Ord(1/\Nf^3)$ \\
{SQED$_3$ ($n=2$)} & $\gamma_{m_\psi}=0.811/\Nf - 0.678/\Nf^2+\Ord(1/\Nf^3)$ \\
{bQED$_3$ ($n=0$)} & $\gamma_{m_\phi}=1.081/\Nf - 5.021/\Nf^2+\Ord(1/\Nf^3)$
\end{tabular}
\vspace*{-0.25cm}
\caption{Numerical mass anomalous dimensions}
\vspace*{-0.25cm}
\label{tab:gammanum}
\end{table}

\section{Dynamical (s)electron mass generation}
\label{sec:massgen}

As an application of our results, we now turn to an estimate of $\Nc$, the critical number of (s)electron flavors which is such that for $\Nf > \Nc$ the (s)electron is massless while for $\Nf < \Nc$ a dynamical mass, with a Miransky-type scaling $m_\text{dyn}\propto\exp(-2\pi/\sqrt{\Nc/\Nf-1})$ \cite{Appelquist:1988sr}, is generated.\footnote{The potentially generated parity-even mass terms are of the form $\mathcal{L}_{\text{dyn}}= m_\psi(\bar{\psi}_i\psi^i-\bar{\psi}_{i+\Nf/2}\psi^{i+\Nf/2}) + m_\phi^2(|\phi_i|^2+|\phi_{i+\Nf/2}|^2)$ with $i=1,...,\Nf/2$ in term of 2-component spinors. Note that only the electron mass term breaks the global flavour symmetry.} 

In the following, we shall only focus on the electron mass generation. Indeed, in the case of bQED$_3$ with $\Nf$ scalars, we did not find any evidence for dynamical scalar mass generation, suggesting that $\Nc=0$ for that model (the situation seems to be different in $4$-dimensions, see \cite{Dagotto:1989gp}). On the other hand, for SQED$_3$ (similarly to the $4$-dimensional case, see \cite{Shamir:1990pa,Shamir:1990pb}) we find a possibility that a selectron mass can be induced by the electron condensate, if the latter exists. As will be seen in the following, our results suggest that electrons do not condense in SQED$_3$. 

Proceeding along the lines of the recent \cite{Metayer:2021rco}, we shall then assume that the electron gap equation for $\mathcal{N}=1$ SQED$_3$ takes the same form as for fQED$_3$, \ie, $\gamma^c_{m_\psi} (1-\gamma^c_{m_\psi})=1/4$, that has to be properly truncated at each order of the $1/\Nf$-expansion and where the mass anomalous dimension $\gamma^c_{m_\psi}$ is given by (\ref{eq:res:gammam}) at $\Nf=\Nc$ (with an all-order estimate of $\gamma_{m_\psi}$, the gap equation reduces to $\gamma^c_{m_\psi}=1/2$). Though semi-phenomenological, such an approach is straightforward and completely gauge invariant. It also takes into account of the feedback of the selectron on the electron that is encoded in $\gamma_{m_\psi}$.

Truncating the gap equation at the LO of the $1/\Nf$ expansion, yields the gauge-invariant value $\Nc=32/\pi^2 = 3.24$ (in terms of 2-component spinors) that coincides with the Landau gauge result of \cite{Koopmans:1989kv}. This LO result suggests that an electron mass is generated for $\Nf=2$ (since $\Nf$ is assumed to be an even integer) thus seemingly breaking both flavour and SUSY symmetries. We find that higher order corrections dramatically change this picture. Indeed, truncating the gap equation at the NLO of the $1/\Nf$ expansion, we find that $\Nc=({16/\pi^2})(1 \pm {\I \,\sqrt{14-\pi^2}/2}) = 1.62\,(1\pm1.02\,\I)$. Such a complex value arises because of the negative NLO contribution (due to the selectron) to the mass anomalous dimension (\ref{eq:res:gammam}), see tab.\ \ref{tab:gammanum}, that prevents the gap equation from having any real valued solution. This calls for a $1/\Nf^3$ computation that is clearly outside the scope of this paper. So, in order to overcome this difficulty, we shall proceed with a resummation of the seemingly alternating asymptotic series. A simple Pad\'e approximant $[1/1]$ of (\ref{eq:res:gammam}) leads~to
\be
\gamma_{m_\psi}=\gamma_{m_\phi}=\frac{8}{28+(\Nf-2)\pi^2}\,.
\ee
Using this new improved value as a input to solve the gap equation non-perturbatively, \ie, $\gamma^c_{m_\psi}=1/2$, yields
\be
\Nc=2(\pi^2-6)/\pi^2 = 0.78\,.
\ee
This result is a strong evidence that, beyond the LO of the $1/\Nf$ expansion, no dynamical (parity-even) mass is generated for the electron in $\mathcal{N}=1$ SQED$_3$. Though a dynamical breaking of SUSY may take place in SQED$_3$ (the Witten index is not well defined with massless matter fields, see, \eg, \cite{Appelquist:1997gq} and references therein), the absence of any electron condensate suggests that SUSY is preserved, in accordance with our perturbative result $\gamma_{m_\psi}=\gamma_{m_\phi}$ found in eq.\ (\ref{eq:res:gammam}).

In closing, let's note that, in the case of fQED$_3$ (for which the gap equation is known exactly up to NLO \cite{Gusynin:2016som,Kotikov:2016prf,Kotikov:2020slw}), the same procedure (this time using (\ref{eq:gammam:qed}) for the mass anomalous dimension) leads at LO to $\Nc=128/(3\pi^2)=4.32$ and at NLO to $\Nc=(64/(3\pi^2))(1+\sqrt{3\pi^2-28}/4)=2.85$ (in terms of 4-component spinors) in accordance with \cite{Gusynin:2016som,Kotikov:2016prf,Kotikov:2020slw}. Although the problem of a complex $\Nc$ is not encountered in this case (because the NLO term in (\ref{eq:gammam:qed}) is positive, see tab.\ \ref{tab:gammanum}), we still provide for completeness the improved $\Nc$ value obtained with resummation, \ie, $\Nc=2(4+3\pi^2)/(3\pi^2)=2.27$. As expected from the effect of radiative corrections, this value is smaller than the exact NLO one but still quite close to it in accordance with the stability of the critical point. In striking contrast with both SQED$_3$ and bQED$_3$, this suggests that a dynamical (flavour breaking and parity-even) mass is indeed generated radiatively for the electron in fQED$_3$ for $\Nf = 1$ and $2$ (in terms of 4-component spinors), or equivalently, for $\Nf =2$ and $4$ (in terms of 2-component spinors). 

\section{Summary and conclusion}
\label{sec:conclusion}

In this paper, we have computed exactly the electron and selectron field and (parity-even) mass anomalous dimensions (as well as the renormalized self energies that we provide for completeness in the appendix) in $\mathcal{N}=1$ SQED$_3$ for an arbitrary covariant gauge fixing and at the NLO of the $1/\Nf$ expansion. Our general framework also allowed us to compute the corresponding anomalous dimensions in bQED$_3$ and recover known results in the case of fQED$_3$. All these quantities correspond to the critical exponents of the considered models at the non-trivial IR fixed point that arises in the large $\Nf$ limit. In the case of SQED$_3$, the (gauge-variant) field anomalous dimensions (\ref{eq:fieldanomalousdims}) were found to be free of $\eps$-scalars; on the other hand, our analysis has shown that $\eps$-scalars play a subtle role in ensuring the equality of the two gauge-invariant mass anomalous dimensions (\ref{eq:res:gammam}). The corresponding results for bQED$_3$ are given by eqs.~(\ref{eq:fieldanomalousdims:bQED}) and (\ref{eq:gammam:bqed}). Note that all these results have a transcendental structure that is similar to that known in the case of fQED$_3$, see (\ref{eq:gammapsinonsusy}) and (\ref{eq:gammam:qed}). There are however noticeable quantitative differences with radiative corrections having a tendency to increase vacuum polarization in bQED$_3$ with respect to fQED$_3$ (see tab.\ \ref{tab:pinum}) while acting oppositely on the mass anomalous dimension (see tab.\ \ref{tab:gammanum}). 
The case of SQED$_3$ is somehow intermediate between fQED$_3$ and bQED$_3$ with, in particular, a tendency of the bosonic contribution from the selectron to reduce the overall electron mass anomalous dimension, due to a smaller LO term and a negative NLO contribution. 
At a non-perturbative level, we also find a marked difference between fQED$_3$ and bQED$_3$. In fQED$_3$, a flavour-breaking parity-invariant mass is generated for (even) $\Nf \leq 4$ (in terms of 2-component spinors) while in bQED$_3$ we do not find any evidence for a dynamically generated scalar mass. In SQED$_3$, the electron condensate, provided it exists, may induce a selectron mass. A resummation of the seemingly alternating series (\ref{eq:res:gammam}) allowed us to evaluate the critical electron flavor number, $\Nc$, that is such that for $\Nf < \Nc$ a dynamical mass for the electron would be generated. The value obtained, $\Nc=0.78$ (in terms of 2-component spinors), strongly suggests that $\mathcal{N}=1$ SQED$_3$ remains in an interacting conformal phase for all values of $\Nf$.

\onecolumn

\section*{Acknowledgments}
We would like to warmly thank J.\ Gracey, V.\ Gusynin, C.\ Herzog  and A.\ Kotikov for inspiring discussions, comments on the manuscript and encouragements. We are grateful to J.\ Gracey for attracting our attention to the case of bQED$_3$ that helped in significantly improving the manuscript, V.\ Gusynin for continuous correspondence and very useful suggestions even in conditions of extreme difficulty, C.\ Herzog for inspiring discussions including on the Witten index and A.\ Kotikov for his meticulous reading of the manuscript.
We thank A.\ James for collaboration at the initial stages of this work. 

\appendix

\section{Renormalized self energies}
\label{sec:app1}

In order to complete our analysis of $\mathcal{N}=1$ SQED$_3$, we provide in this Appendix the exact expressions of the renormalized
matter self-energies at $\Ord(1/\Nf^2)$.
They can be derived from the relations:
\be
\Sigma_{pr}^{\psi}=1-(1-\Sigma_p^\psi)Z_\psi\,,\qquad
\Sigma_{mr}^{\psi}=1-(1+\Sigma_m^\psi)Z_\psi Z_{m_\psi}\,,\qquad
\Sigma_{pr}^{\phi}=1-(1-\Sigma_p^\phi)Z_\phi\,,\qquad
\Sigma_{mr}^{\phi}=1-(1+\Sigma_m^\phi)Z_\phi Z_{m_\phi}^2\,, \nonumber
\ee
that require the computation of the finite parts, \ie, of $\Ord(\eps^0)$, of the (bare) matter self energies and from which one can straightforwardly recover the corresponding renormalized matter propagators. All calculations done, exactly up to $\Ord(1/\Nf^2)$, the renormalized matter self-energies of $\mathcal{N}=1$ SQED$_3$ ($S=\mathcal{E}=1$) with $n=2$-component spinors read
\begin{subequations}
\begin{flalign}
&\Sigma_{pr}^\psi=-\frac{2}{\pi^2\Nf}\Big[2+(2-\Lpt)\xi\Big]+\frac{8}{\pi^4\Nf^2}\Big[3+10\xi+\Big(1-(5-\xi)\xi-\tfrac{1}{4}\xi^2\Lpt\Big)\Lpt-\Big(4+\tfrac{1}{2}(12+\xi)\xi-3\xi\Lpt\Big)\zeta_2\Big]+\Ord(\eps)\,,\nonumber\\
&\Sigma_{mr}^\psi=-\frac{2}{\pi^2\Nf}\Big[(2+\xi)(4-\Lpt)\Big]+\frac{8}{\pi^4\Nf^2}\Big[37-16 C_4+2(4-\xi)\xi-\Big(5-2(1+\xi)\xi+\tfrac{1}{4}(2+\xi)^2\Lpt\Big)\Lpt\nonumber\\[-9pt]
& \hspace{10.43cm}-\Big(18+4C_2+\tfrac{1}{2}(28+\xi)\xi-3(2+\xi)\Lpt\Big)\zeta_2\Big]+\Ord(\eps)\,,\nonumber\\[-3pt] 
&\Sigma_{pr}^\phi=\frac{2}{\pi^2\Nf}\Big[4-(2-\xi)\Lpt\Big]-\frac{8}{\pi^4\Nf^2}\Big[35-8C_4-\Big(17-8\xi-\tfrac{1}{4}(2-\xi)^2\Lpt\Big)\Lpt-\Big(9+2C_2+\tfrac{1}{2}(4-\xi)\xi-3(2-\xi)\Lpt\Big)\zeta_2\Big]+\Ord(\eps)\,,\nonumber\\
&\Sigma_{mr}^\phi=-\frac{2}{\pi^2\Nf}\Big[14-3\xi-(2+\xi)\Lpt\Big]+\frac{4}{\pi^4\Nf^2}\Big[123-46C_4-2(12+\xi)\xi-\Big(2+(4+3\xi)\xi+\tfrac{1}{2}(2+\xi)^2\Lpt\Big)\Lpt\nonumber\\[-9pt]
& \hspace{9.7cm}-\tfrac{1}{4}\Big(227+46C_2-4(14-\xi)\xi-24(2+\xi)\Lpt\Big)\zeta_2\Big]+\Ord(\eps)\,,\nonumber
\end{flalign}
\end{subequations}
with $\Lpt=\log(-4p^2/\bar{\mu}^2)$ and $\zeta_2=\pi^2/6$, as well as $C_2=\text{CL}_2(\pi/2)=0.916$ the Catalan number and $C_4=\text{CL}_4(\pi/2)=0.989$ where $\text{CL}_n(z)$ is the Clausen function. For completeness, we also provide the bQED$_3$ ($n=\EP=0$, $S=1$) case
\begin{subequations}
\begin{flalign}
&\Sigma_{pr}^{\phi(\text{b})}=\frac{4}{9\pi^2\Nf}\Big[56-3(8-3\xi)\Lpt\Big]-\frac{32}{81\pi^4\Nf^2}\Big[5455-648C_4-3\Big(884-330\xi-\tfrac{3}{4}(8-3\xi)^2\Lpt\Big)\Lpt\nonumber\\[-7pt] 
& \hspace{10cm}-27\Big(39+6C_2+\tfrac{1}{2}(16-3\xi)\xi-3(8-3\xi)\Lpt\Big)\zeta_2\Big]+\Ord(\eps)\,,\nonumber\\[-5pt] 
&\Sigma_{mr}^{\phi(\text{b})}=-\frac{4}{\pi^2\Nf}\Big[8-3\xi-\xi\Lpt\Big]+\frac{16}{9\pi^4\Nf^2}\Big[1513-630C_4-6(86+3\xi)\xi-\Big(4(18+23\xi)+\tfrac{9}{2}(6+\Lpt)\xi^2\Big)\Lpt\nonumber\\[-5pt] 
& \hspace{10.6cm} -\tfrac{3}{4}\Big(365+210C_2-72(3+\Lpt)\xi+12\xi^2\Big)\zeta_2\Big]+\Ord(\eps)\,,\nonumber
\end{flalign}
\end{subequations}
as well as the non-SUSY ($S=0$) fQED$_3$ case with $n=4$-component electrons
\begin{subequations}
\begin{flalign}
&\Sigma_{pr}^{\psi(\text{f})}=\frac{2}{9\pi^2\Nf}\Big[2+3(2-3\xi)(2-\Lpt)\Big]-\frac{8}{81\pi^4\Nf^2}\Big[787-846\xi-3\Big(110-3(59-9\xi)\xi-\tfrac{3}{4}(2-3\xi)^2\Lpt\Big)\Lpt\nonumber\\[-7pt] 
& \hspace{11cm}-27\Big(16-\tfrac{1}{2}(32+3\xi)\xi-3(2-3\xi)\Lpt\Big)\zeta_2\Big]+\Ord(\eps)\,,\nonumber\\[-5pt] 
&\Sigma_{mr}^{\psi(\text{f})}=-\frac{2}{\pi^2\Nf}\Big[6+4\xi-(2+\xi)\Lpt\Big]+\frac{8}{9\pi^4\Nf^2}\Big[5(23-36C_4)+2(26-9\xi)\xi-\Big(26-(17+18\xi)\xi+\tfrac{9}{4}(2+\xi)^2\Lpt\Big)\Lpt\nonumber\\[-5pt] 
& \hspace{10.6cm} -\tfrac{9}{2}\Big(17+10C_2+(28+\xi)\xi-6(2+\xi)\Lpt\Big)\zeta_2\Big]+\Ord(\eps)\,.\nonumber
\end{flalign}
\end{subequations}

\begin{multicols}{2}

\bibliographystyle{custom}
\bibliography{sqed3m.bib}

\end{multicols}

%\end{fmffile}
\end{document}